\documentclass[onecolumn,amsmath,amssymb,superscriptaddress,nofootinbib,11pt,a4paper]{revtex4}

\pdfoutput=1
\usepackage[T1]{fontenc}
\usepackage{CJK}
\usepackage{xcolor}
\usepackage{amsfonts}
\usepackage{epstopdf}
\usepackage{wrapfig} 
\usepackage{subcaption}
\usepackage{graphicx}  
\usepackage{dcolumn}   
\usepackage{bm}
\usepackage{float}
\usepackage[T1]{fontenc}
\usepackage{CJK}
\usepackage{xcolor}
\usepackage{amsfonts}
\usepackage{epstopdf}
\usepackage{wrapfig} 
\usepackage{subcaption} 
\usepackage{graphicx}  
\usepackage{dcolumn}   
\usepackage{bm}
\usepackage{float}
\usepackage[utf8]{inputenc}
\usepackage[T1]{fontenc}

\begin{document}

\title{Hotspot Images from Magnetic Reconnection Processes in the plunging Region of a Kerr Black Hole}

\author{Xiao-Xiong Zeng}
\email{xxzengphysics@163.com}
\affiliation{College of Physics and Optoelectronic Engineering, Chongqing Normal University, \\Chongqing 401331, China}

\author{Yun Hong}
\email{hongyun1968@163.com}
\affiliation{Department of physics, college of computer science and electronic information engineering,\\
Chongqing Technology and Business University, Chongqing 400067, China}

\author{Ke Wang\footnote{Electronic address: kkwwang2025@163.com  (Corresponding author)}}
\affiliation{School of Material Science and Engineering, Chongqing Jiaotong University, \\Chongqing 400074, China}

\begin{abstract}
{Employing the hotspot imaging technique, this work investigates the plasma motion trajectories prior to and following the Comisso-Asenjo mechanism within the plunging region. After a concise overview of the magnetic reconnection process in the plunging region of a Kerr black hole, we present the hotspot model and the associated imaging methodology. Through numerical simulations, we separately examine the hotspot images in the plunging region under three conditions: no magnetic reconnection, with magnetic reconnection, and when the escape condition fails. These outcomes are also contrasted with hotspot images in the circular orbit zone. Our findings reveal that for hotspot images without magnetic reconnection, when the plasma follows plunging orbits, the flare strength gradually declines; conversely, for circular orbits, the flare strength remains approximately constant. Additionally, we observe that the signal indicative of energy extraction is less conspicuous in the plunging region compared to the circular orbit region.}
\end{abstract}

\maketitle
 \newpage
\section{Introduction}
Over recent years, extensive observational evidence has confirmed the existence of a supermassive black hole at the center of our Milky Way \cite{12,13}. Near-infrared flare events originating from the vicinity of the Galactic center black hole's event horizon have also drawn considerable attention \cite{14}. To date, numerous investigations have concentrated on black hole imagery and flare phenomena, with a major research direction being the mechanism that drives black hole accretion flows to produce near-infrared flares \cite{15}. Several approaches exist for studying sources capable of generating such near-infrared flares; one effective method involves computing hotspots orbiting the central black hole \cite{16,17,18,19,20,21,22}. In this approach, a semi-analytical hotspot imaging model is typically employed, and numerical techniques are used to compute photon trajectories.

Meanwhile, research on magnetic reconnection within the framework of general relativity has advanced rapidly in recent years \cite{23,24,25,26}. Magnetic reconnection can account for a variety of astrophysical phenomena, including stellar flares, coronal mass ejections, and gamma-ray burst jets \cite{27}. Furthermore, magnetic reconnection events can accelerate magnetized fluids to near-light speeds, thereby serving as a trigger for the Penrose process \cite{28}, which enables the extraction of rotational energy from the black hole. The method proposed by Comisso and Asenjo for extracting black hole energy via magnetic reconnection \cite{2} has been extensively generalized to various gravitational backgrounds \cite{29,30,31,32,33,34}. In particular, Ref. \cite{1} extended the Comisso-Asenjo process to the plunging region. Ref. \cite{38} was the first to demonstrate that, under certain conditions, the energy extraction efficiency in the plunging region can surpass that in the circular orbit region. Utilizing the Comisso-Asenjo process to extract black hole energy from the plunging region has also been extended to diverse black hole spacetimes \cite{35,36,37,40}. Notably, Ref. \cite{39} proved that this plasma-driven Penrose process is energetically feasible under realistic astrophysical conditions, thereby enhancing the likelihood of its actual occurrence. Nevertheless, relatively few studies have addressed hotspot imaging associated with the Penrose process driven by the Comisso-Asenjo mechanism. Ref. \cite{10} studied the hotspot imaging of this process in a Kerr black hole, i.e., observing the plasma motion during the process, and found that it could potentially produce three flares, with the first flare possibly serving as a signature of Penrose process energy extraction. Recently, we extended this study to Kerr-Sen spacetime \cite{11}, Kerr-AdS spacetime \cite{42}, and rotating non-commutative spacetime \cite{43}. Not only did we discover similar phenomena, but we also found that expansion parameters, the cosmological constant, and the non-commutative parameter significantly affect hotspot imaging. However, the hotspot imaging studied in the aforementioned works was conducted exclusively in the circular orbit region. In this paper, we extend the work of Ref. \cite{10} to the plunging region. We will discuss the characteristics of hotspot imaging in the plunging region and how they differ from those in the circular orbit region. Our results show that for hotspot images without magnetic reconnection, when the plasma follows a plunging orbit, the flare intensity gradually decreases; whereas for circular orbits, the flare intensity remains nearly unchanged. Additionally, in the plunging region, the signal for identifying energy extraction is weaker than that in the circular orbit region. To identify the energy extraction signal, it is more effective to conduct studies in the circular orbit region, although the signal can still be identified in the plunging region.

The remainder of this paper is organized as follows. Section 2 introduces the magnetic reconnection process in the plunging region of a Kerr black hole. Section 3 describes the hotspot model and imaging method. Section 4 presents the numerical observational results. A summary is provided in Section 5. Throughout this paper, we adopt natural units ($c=G=1$).

\section{Brief Overview of Magnetic Reconnection in the plunging Region of a Kerr Black Hole}

In this section, we briefly review the magnetic reconnection process in the plunging region as presented in Ref. \cite{1}, also known as the Comisso-Asenjo process, originally introduced in Ref. \cite{2}. In natural units and Boyer-Lindquist (BL) coordinates, the Kerr metric is given by \cite{3}
\begin{equation}
ds^{2} = -(1-\frac{2Mr}{\Sigma})dt^{2}+\frac{\Sigma}{\Delta}dr^{2}+\Sigma d \theta^{2}-\frac{4aMr}{\Sigma}\sin^{2}\theta dtd\phi+\sin^{2}\theta\left(r^2+a^{2}+\frac{2Mra^{2}\sin^{2}\theta }{\Sigma}\right)d\phi^{2},    
\end{equation}
where
\begin{equation}
\Sigma=r^2+a^{2}\cos^{2}\theta,\Delta=r^2+a^{2}-2Mr,  
\end{equation}
with $M$ being the black hole mass and $a$ the spin parameter. The outer horizon is located at
\begin{equation}
r_h = M  + \sqrt{ M^2 - a^2}.  
\end{equation}
The outer ergosphere boundary is given by
\begin{equation}
r_E = M +\sqrt{M^2-  a^2 \cos^2 \theta}.   
\end{equation}
The innermost stable circular orbit (ISCO) is at \cite{44}
\begin{equation}
r_{I} = M \left[ 3 + Z_2 \mp  \sqrt{ (3 - Z_1)(3 + Z_1 + 2Z_2) } \right] ,  
\end{equation}
where
\begin{equation}
Z_1 = 1 + \sqrt[3]{1 - \frac{a^2}{M^2}}\left( \sqrt[3]{1 + \frac{a}{M}} + \sqrt[3]{1 - \frac{a}{M}} \right),
Z_2 = \sqrt{3\frac{a^2}{M^2} + Z_1^2},
\end{equation}
with the minus and plus signs corresponding to prograde and retrograde orbits, respectively. In the plunging region, only prograde orbits need be considered for energy extraction (see footnote 2 of Ref. \cite{1}). We set $M=1$ throughout.

The spacetime metric can be decomposed in the 3+1 form
\begin{equation}
ds^{2}=-\alpha^{2}dt^{2}+\sum_{i=1}^{3}\bigl{(}\sqrt{g_{ii}}dx^{i}-\alpha\beta^{i}dt\bigr{)}^{2}=g_{\mu\nu}dx^{\mu}dx^{\nu},   
\end{equation}
where
\begin{equation}
\omega^{\phi}=-g_{t\phi}/g_{\phi\phi},\quad \alpha=\sqrt{-g_{tt}+\frac{g_{t\phi}^{2}}{g_{\phi\phi}}},\quad \beta^{i}=\delta_{i\phi}\frac{\sqrt{g_{\phi\phi}}\,\omega^{\phi}} {\alpha}.  
\end{equation}
Using the Zero Angular Momentum Observer (ZAMO) tetrad
\begin{equation}
\hat{e}_0 = \frac{1}{\alpha}(\partial_t + \omega^\phi \partial_\phi),\quad \hat{e}_1 = \frac{1}{\sqrt{g_{rr}}} \partial_r,\quad \hat{e}_2 = \frac{1}{\sqrt{g_{\theta\theta}}} \partial_\theta,\quad \hat{e}_3 = \frac{1}{\sqrt{g_{\phi\phi}}} \partial_\phi. 
\end{equation}
According to the simplifying assumptions of Ref. \cite{2}, the current sheet is confined to the equatorial plane. The transformation between the four-velocity in the BL frame and the ZAMO frame is
\begin{equation}
\hat{U}^{\mu}=\hat{\gamma}\left[1, \hat{v}^{(r)}, 0, \hat{v}^{(\phi)}\right]=\left[\frac{E-\omega^{\phi} L}{\alpha}, \sqrt{g_{r r}} U^{r}, 0, \frac{L}{\sqrt{g_{\phi \phi}}}\right],\label{10}
\end{equation}
where
\begin{equation}
\hat{v}=\sqrt{\left(\hat{v}^{(r)}\right)^{2}+\left(\hat{v}^{(\phi)}\right)^{2}},\quad \hat{\gamma}=1/\sqrt{(1-\hat{v}^{2})}.
\end{equation}
In the plunging region, the energy $E$ and angular momentum $L$ of the current sheet take the ISCO values
\begin{equation}
E_I=\frac{a+r_I^{3/2}-2 \sqrt{r_I}}{\sqrt{2 a r_I^{3/2}+r_I^3-3 r_I^2}},\quad L_I=\frac{a^2+r_I^2-2 a\sqrt{r_I}}{\sqrt{2 a r_I^{3/2}+r_I^3-3 r_I^2}}.\label{12}
\end{equation}
The radial four-velocity of the current sheet is given by \cite{4}
\begin{equation}
U^{r}_K=-\sqrt{\frac{2}{3r_{ I}}}\left(\frac{r_{I}}{r}-1\right)^{3/2}.\label{13}
\end{equation}
From \eqref{10}, the three-velocity components and Lorentz factor in the ZAMO frame are
\begin{equation}
\hat{\gamma}_K=\frac{E_I-L_I \omega^{\phi}}{\alpha},\quad \hat{v}^{(r)}_K=\sqrt{g_{r r}} U^{r}_K/\hat{\gamma}_K,\quad \hat{v}^{(\phi)}_K=L_I/(\sqrt{g_{\phi\phi}} \hat{\gamma}_K),\quad \hat{v}_K=\sqrt{\left(\hat{v}_K^{(r)}\right)^{2}+\left(\hat{v}_K^{(\phi)}\right)^{2}}.
\end{equation}
The fluid rest frame is obtained via a tetrad transformation
\begin{equation}
\begin{aligned}
\sigma_{0}&=\hat{\gamma}_{K}\left[\hat{e}_0+\hat{v}^{(r)}_{K}\hat{e}_{1}+\hat{v}^{({\phi})}_{K}\hat{e}_{3}\right],\\
\sigma_{1} &=\frac{1}{\hat{v}_{K}}\left[\hat{v}^{({\phi})}_{K}\hat{e}_{1}-\hat{v}^{(r)}_{K}\hat{e}_{3}\right],\quad \sigma_{2}=\hat{e}_{2}, \\
\sigma_{3}&=\hat{\gamma}_{K}\left[\hat{v}_{K}\hat{e}_{0}+\frac{\hat{v} ^{(r)}_{K}}{\hat{v}_{K}}\hat{e}_{1}+\frac{\hat{v}^{({\phi})}_{K}}{ \hat{v}_{K}}\hat{e}_{3}\right].
\end{aligned}
\end{equation}
Let $\xi$ denote the azimuthal angle of the magnetic field in the fluid rest frame. The four-velocities of the accelerated and decelerated plasma ejected from the reconnection layer in that frame are
\begin{equation}
u^{\prime\mu} = \gamma_{out}[\sigma_{0} \pm v_{out}(\cos\xi\sigma_{3}+ \sin\xi\sigma_{1})],   
\end{equation}
where $+$ and $-$ label the accelerated and decelerated plasma, respectively. Here $v_{out}$ is the outflow speed in the fluid rest frame, and $\gamma_{out}$ is the corresponding Lorentz factor, satisfying \cite{2}
\begin{equation}
v_{ out }=\sqrt{\frac{\sigma}{1+\sigma}},\qquad \gamma_{out}= (1-v_{ out }^{2})^{-1 / 2},
\end{equation}
with $\sigma$ the upstream magnetization parameter. In the ZAMO frame, the four-velocities become
\begin{equation}
\begin{aligned}
\hat{u}^{\mu} &= \gamma_{out}\left[\hat{\gamma}_{K}[\hat{e}_0+\hat{v}^{(r)}_{K}\hat{e}_{1}+\hat{v}^{({\phi})}_{K}\hat{e}_{3}] \right. \\
&\quad \left. \pm  v_{out}\left(\cos\xi(\hat{\gamma}_{K}[\hat{v}_{K}\hat{e}_{0}+\frac{\hat{v} ^{(r)}_{K}}{\hat{v}_{K}}\hat{e}_{1}+\frac{\hat{v}^{({\phi})}_{K}}{ \hat{v}_{K}}\hat{e}_{3}]) + 
\sin\xi(\frac{1}{\hat{v}_{K}}[\hat{v}^{({\phi})}_{K}\hat{e}_{1}-\hat{v}^{(r)}_{K}\hat{e}_{3}])\right)\right] \\
&= \hat{\gamma}_{out}(1,\hat{v}^{r},0,\hat{v}^{\phi}), 
\end{aligned}
\end{equation}
where
\begin{equation}
\begin{aligned}
\hat{\gamma}_{out} &= \gamma_{out}\hat{\gamma}_{K}(1 \pm \hat{v}_{K}v_{out}\cos\xi), \\
\hat{v}^{r} &= \frac{ \hat{\gamma}_{K} \hat{v}^{(r)}_{K} \pm v_{out} \cos\xi \hat{\gamma}_{K} \frac{\hat{v}^{(r)}_{K}}{\hat{v}_{K}} \pm v_{out} \sin\xi \frac{1}{\hat{v}_{K}} \hat{v}^{(\phi)}_{K} }{ \hat{\gamma}_{K} (1 \pm v_{out} \cos\xi \hat{v}_{K}) },\\
\hat{v}^{\phi} &= \frac{ \hat{\gamma}_{K} \hat{v}^{(\phi)}_{K} \pm v_{out} \cos\xi \hat{\gamma}_{K} \frac{\hat{v}^{(\phi)}_{K}}{\hat{v}_{K}} \mp v_{out} \sin\xi \frac{\hat{v}^{(r)}_{K}}{\hat{v}_{K}} }{ \hat{\gamma}_{K} (1 \pm v_{out} \cos\xi \hat{v}_{K}) }.
\end{aligned}
\end{equation}
Using \eqref{10}, the energy, angular momentum, and radial momentum in the BL frame are
\begin{equation}
p_r= \hat{\gamma}_{out}\hat{v}^{r}  \sqrt{g_{rr}},\quad \mathcal{L}=\sqrt{g_{\phi \phi}}\hat{\gamma}_{out}\hat{v}^{\phi}=p_\phi,\quad \mathcal{E}=\alpha\hat{\gamma}_{out}+\omega^\phi \mathcal{L}=-p_t.\label{20}
\end{equation}
Equations \eqref{20} govern the motion of the accelerated and decelerated plasma and are essential for hotspot imaging. If no magnetic reconnection occurs, the hotspot imaging is simulated numerically using \eqref{12} and \eqref{13}. Assuming highly efficient reconnection where magnetic energy is almost entirely converted into plasma kinetic energy, the electromagnetic field energy can be neglected relative to the fluid energy. Treating the plasma as an adiabatic incompressible fluid, the energy-at-infinity per unit enthalpy for the accelerated and decelerated plasma is approximated as \cite{1,2}
\begin{equation}
\begin{aligned}
\varepsilon_{ \pm}&=\alpha \hat{\gamma}_{K} \gamma_{{out }}\left[\left(1+\beta^{\phi} \hat{v}_{K}^{(\phi)}\right) \pm v_{ {out }}\left(\hat{v}_{K}+\beta^{\phi} \frac{\hat{v}_{K}^{(\phi)}}{\hat{v}_{K}}\right) \cos \xi \mp v_{ {out }} \beta^{\phi} \frac{\hat{v}_{K}^{(r)}}{\hat{\gamma}_{K} \hat{v}_{K}} \sin \xi\right]\\&-\alpha \left( 4 \hat{\gamma}_{K} \gamma_{{out}} \left(1 \pm \hat{v}_{K} v_{{out}} \cos \xi\right) \right)^{-1}.
\end{aligned}
\end{equation}
For energy extraction, the following conditions must hold \cite{2}
\begin{equation}
  \varepsilon_{+}>0,\quad \varepsilon_{-}<0.\label{22}
\end{equation}
We assume a relativistically hot plasma with a polytropic index of $4/3$. In the plunging region, an additional requirement is the escape condition: even if \eqref{22} is satisfied, $\varepsilon_{+}$ may still fail to escape to infinity and instead fall into the event horizon, preventing energy extraction. The radial motion can be described by an effective potential
\begin{equation}
V_{eff}= g_{rr}u_{r}^{2}+g_{\theta\theta}u_{\theta}^{2}=g_{rr}p_r^{2}=\frac{\mathcal{E}^{2}g_{\phi\phi}+2\mathcal{E}\mathcal{L}g_{t\phi}+\mathcal{L}^{2}g_{tt}}{g_{t\phi}^{2}-g_ {tt}g_{\phi\phi}}-1.
\end{equation}
The second equality uses the equatorial plane condition. For given $\sigma$ and $\xi$, $\mathcal{E}$ and $\mathcal{L}$ are functions of the X-point radius $r_X$
\begin{equation}
\mathcal{E}=\mathcal{E}(r_X),\quad \mathcal{L}=\mathcal{L}(r_X). \label{24}
\end{equation}
Equatorial current sheets are susceptible to the plasmoid instability \cite{8,9}, which fragments them into multiple X-points. The dominant X-point, located at the separatrix intersection that encloses the global reconnection outflow, governs the overall reconnection dynamics. From \eqref{20}, the $p_r$ of $\varepsilon_{+}$ is always inward. For $\varepsilon_{+}$ to escape, the potential at $r=r_X$ must be sufficiently large to reverse its direction. The potential typically has an extremum at $r=r_e$, which depends on $r_X$ via \eqref{24}. Hence the escape condition for $\varepsilon_{+}$ is
\begin{equation}
V(r_{e}(r_{X}),\,\mathcal{E}(r_{X}),\,\mathcal{L}(r_{X}\,))<0,\quad \text{if } r_{h}<r_{e}(r_{X})<r_{X}.\label{25}
\end{equation}
Equations \eqref{22} and \eqref{25} constitute the energy extraction conditions in the plunging region. For hotspot imaging, only \eqref{12}, \eqref{13}, and \eqref{20} are required.

\section{Hotspot Model and Imaging Technique}

In this section, we describe the hotspot model and the imaging methodology. Following Ref. \cite{10}, our goal is not to study the influence of radiation mechanisms on imaging or flare phenomena; thus the model does not incorporate detailed emission spectra. We assume isotropic, frequency-independent emission, effectively modeling the hotspot as a broadband source with a flat spectrum. The plasma is treated as a transparent hotspot whose emissivity follows a Gaussian distribution
\begin{equation}
J = e^{-\frac{1}{2}\left(\frac{x^2}{s^2}\right)},
\end{equation}
where $J$ is the emissivity, $s$ the Gaussian width, and $x$ the distance from the hotspot center. To image the moving hotspot, we must trace the source trajectory and its radiative transfer. According to \eqref{20}, the trajectory is obtained by numerically integrating the geodesic equations in Hamilton–Jacobi form. For radiative transfer, we employ a backward ray-tracing method combined with a fisheye camera model; the computational details are given in Appendix B of Ref. \cite{5}. The camera is placed in the ZAMO frame, with its tetrad
\begin{equation}
\begin{aligned}
\hat{e}_{(t)} &= \left(g_{\phi\phi}\partial_t - g_{\phi t}\partial_\phi\right) \left[g_{\phi\phi}\left(g_{\phi t}^2 - g_{\phi\phi}g_{tt}\right)\right]^{-1/2},\quad \hat{e}_{(r)} = -\partial_r \left(g_{rr}\right)^{-1/2},\\
\hat{e}_{(\theta)} &= \partial_\theta \left(g_{\theta\theta}\right)^{-1/2},\quad \hat{e}_{(\phi)} = -\partial_\phi \left(g_{\phi\phi}\right)^{-1/2}.
\end{aligned}
\end{equation}
The intensity on the image plane is governed by the radiative transfer equation \cite{6,Zeng:2025kqw,Zeng:2021dlj,Zeng:2021mok}
\begin{equation}
\frac{d}{d\lambda}\left(I_{\nu}\,\nu^{-3}\right)=J_{\nu}\,\nu^{-2},
\end{equation}
where $\lambda$ is the affine parameter along the null geodesic, and $I_{\nu}$ and $J_{\nu}$ are the specific intensity and emissivity at frequency $\nu$. Absorption is neglected. Our numerical scheme produces a series of snapshots, with the photon arrival time equal to the orbital time plus the light travel time from the hotspot to the observer. The flux centroid position on the camera plane is determined for each frame. According to the camera definition in Ref. \cite{5}, the flux on the $(i,j)$-th pixel is \cite{7}
\begin{equation}
F(i,j)=I_{o}S\cos\left(2\arctan\left(n^{-1}\tan\left(\frac{\alpha_{fov}}{2}\right)\left(\left(i-\frac{n+1}{2}\right)^{2}+\left(j-\frac{n+1}{2}\right)^{2}\right)^{1/2}\right)\right),
\end{equation}
where $S$ is the area of a single pixel, $n$ the number of pixels per dimension ($i,j=1,\dots,n$), and $\alpha_{fov}$ the camera's field of view. The centroid position $\vec{x}(t)$ of each image is then
\begin{equation}
\vec{x}(t)=\left(\sum_{i,j}\vec{x}(i,j)F(i,j)\right)\left(\sum_{i,j}F(i,j)\right)^{-1},
\end{equation}
with $\vec{x}(i,j)$ the spatial coordinates of pixel $(i,j)$. The total flux in a snapshot, $\sum_{i,j}F(i,j)$, is the flux associated with $\vec{x}_c(t)$. This framework allows us to track the temporal evolution of the brightness centroid and its flux.

\section{Observational Results}

In this section, we present the hotspot imaging results for the plunging region. The parameters are set as $\sigma=20$, $\xi=\pi/12$, $s=0.2$. The observer's azimuthal angle is $\phi_0=\pi/2$, inclination $\theta_0=\pi/10$, and radial distance 200. The time when the hotspot first appears on the screen is $t=0$. We consider two scenarios. First, $a=0.94$, $r_X=1.6$ (X-point radius). The ergosphere lies at $r\in(1.34,2)$, ISCO at $2.02$, critical radius $r_e$ for escape at $1.52$, and $\varepsilon_{-}=-0.54$, $\varepsilon_{+}=8.01$, satisfying energy extraction. Second, $a=0.99$, $r_X=1.3$ (near-extremal case). The ergosphere is at $r\in(1.14,2)$, ISCO at $1.45$, $r_e=1.21$, and $\varepsilon_{-}=-1.25$, $\varepsilon_{+}=7.97$, also satisfying energy extraction. Due to differences in energy extraction conditions, $r_X$ cannot be identical for both spins.

\subsection{$a=0.94$, $r_X=1.6$}

First, we show hotspot images without magnetic reconnection, where the current sheet plunges from ISCO to the event horizon.

\begin{figure}[!h]
  \centering
  \begin{subfigure}{0.26\textwidth}
    \centering
    \includegraphics[width=\linewidth]{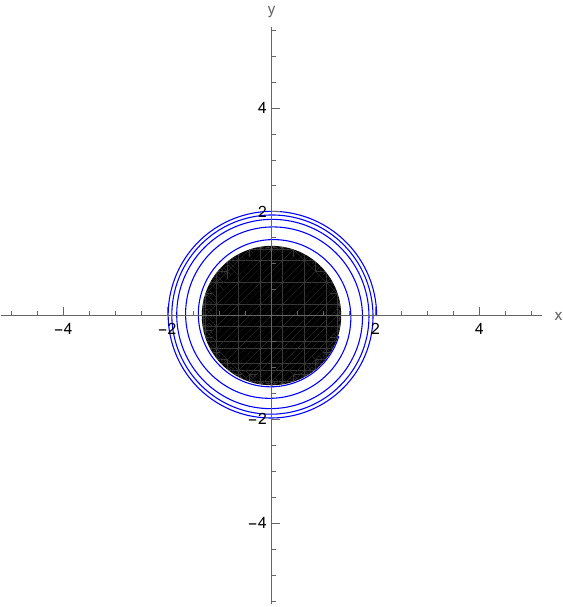}
    \label{fig:1a}
  \end{subfigure}
  \begin{subfigure}{0.26\textwidth}
    \centering
    \includegraphics[width=\linewidth]{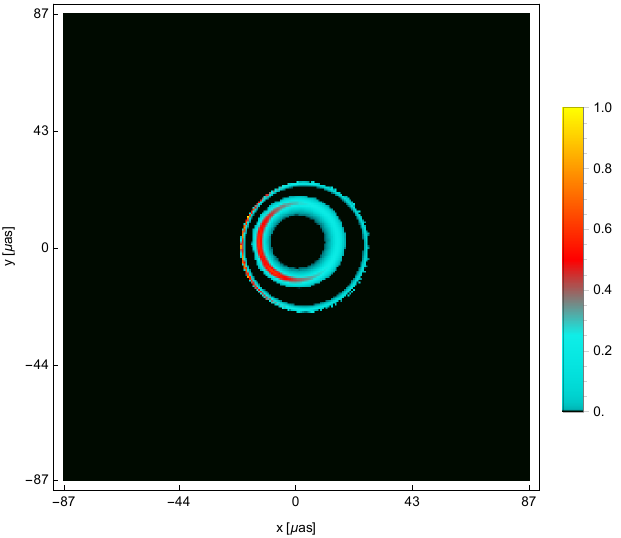} 
    \label{fig:1b}  
  \end{subfigure}
  \begin{subfigure}{0.26\textwidth}
    \centering
    \includegraphics[width=\linewidth]{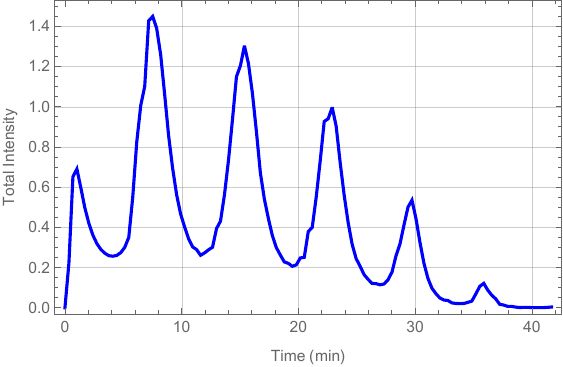} 
    \label{fig:1c} 
  \end{subfigure}
  \caption{For $a=0.94$ in the plunging region without magnetic reconnection: left panel shows the current sheet trajectory in a 2D Cartesian coordinate system; middle panel shows the normalized intensity distribution as seen by the observer (time-averaged radiation intensity, normalized by $I/I_{MAX}$; outer ring indicates secondary/higher-order images); right panel shows the light curve (total flux vs. observation time).}
  \label{fig:1}
\end{figure}

From Fig. \ref{fig:1}, as the current sheet plunges from ISCO to the horizon, two rings appear: the inner ring is the primary image, the outer ring secondary/higher-order images. The light curve exhibits six flares, with each flare after the first gradually weakening. This occurs because the motion approximates Keplerian orbits with decreasing radius; as the orbital radius shrinks, flares become progressively weaker. In Ref. \cite{10}, for over two Keplerian orbits at constant radius\footnote{The magnetization parameter and magnetic field azimuthal angle they used differ from those in this paper, but the absence of magnetic reconnection is independent of these two parameters.}, two prominent flares of similar intensity were seen. Our image corresponds to nearly five decaying-radius orbits, yielding five diminishing flares. The first flare corresponds to the initial bump in Ref. \cite{10}, but here it is more pronounced.

\begin{figure}[!h]
  \centering
  \begin{subfigure}{0.22\textwidth}
    \centering
    \includegraphics[width=\linewidth]{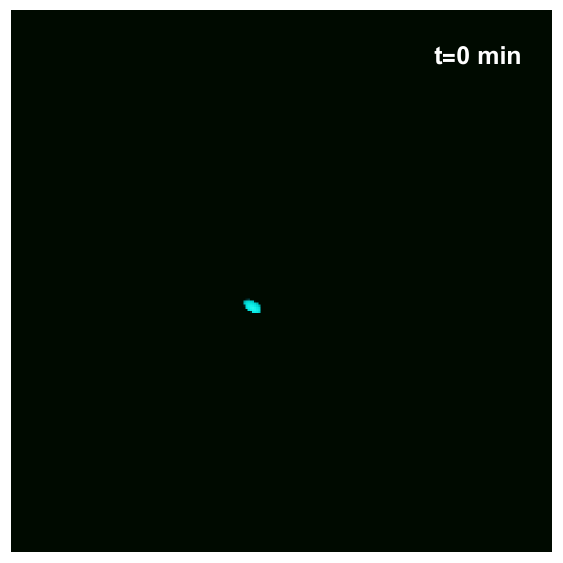} 
  \end{subfigure}
  \begin{subfigure}{0.22\textwidth}
    \centering
    \includegraphics[width=\linewidth]{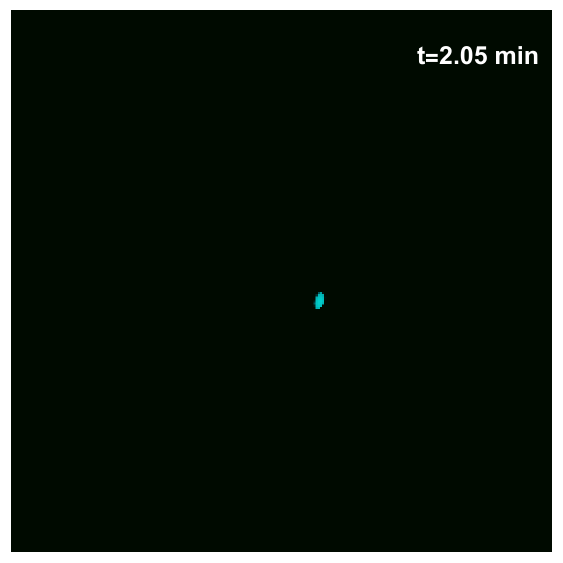} 
  \end{subfigure}
 \begin{subfigure}{0.22\textwidth}
    \centering
    \includegraphics[width=\linewidth]{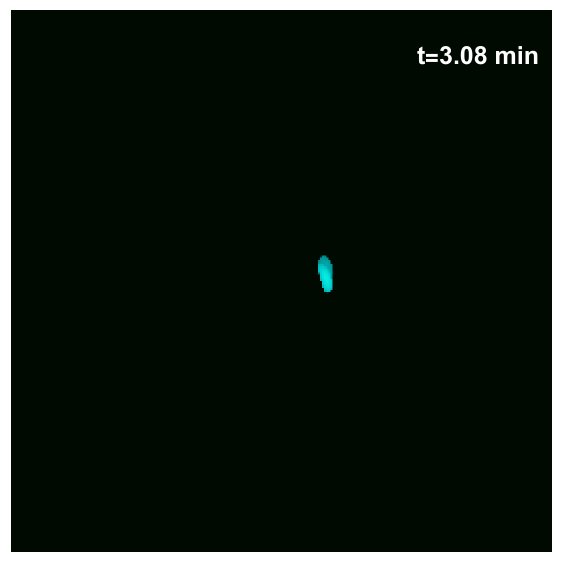}  
  \end{subfigure}
 \begin{subfigure}{0.22\textwidth}
    \centering
    \includegraphics[width=\linewidth]{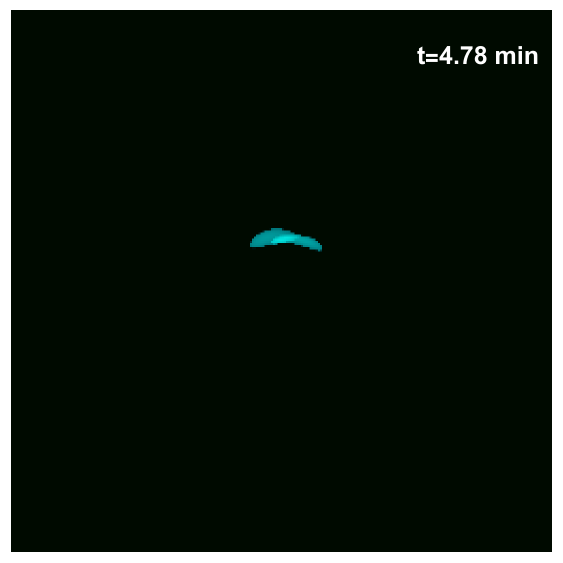} 
  \end{subfigure}
 \begin{subfigure}{0.22\textwidth}
    \centering
    \includegraphics[width=\linewidth]{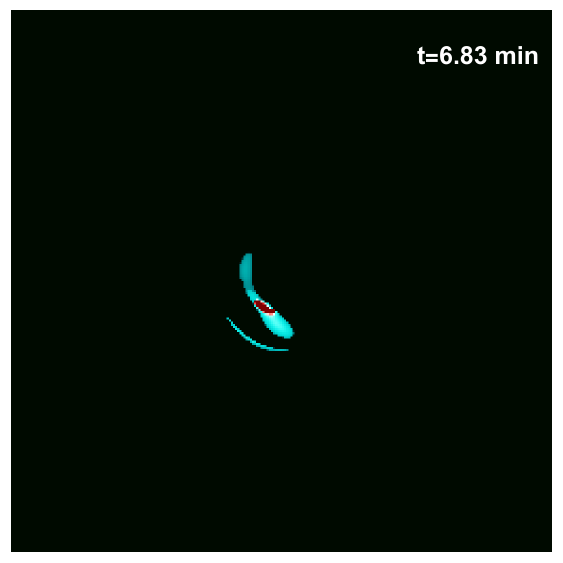}  
  \end{subfigure}
 \begin{subfigure}{0.22\textwidth}
    \centering
    \includegraphics[width=\linewidth]{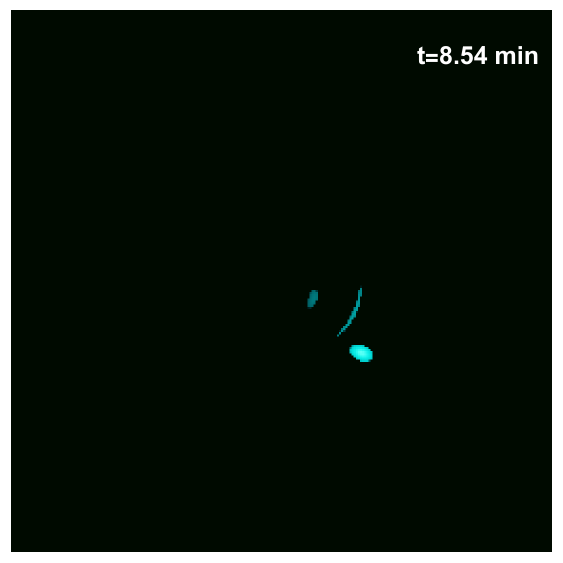} 
  \end{subfigure} 
 \begin{subfigure}{0.22\textwidth}
    \centering
    \includegraphics[width=\linewidth]{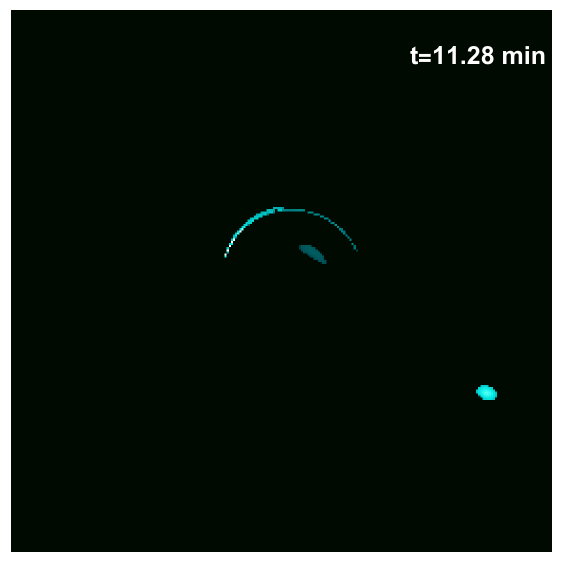} 
  \end{subfigure} 
  \begin{subfigure}{0.22\textwidth}
    \centering
    \includegraphics[width=\linewidth]{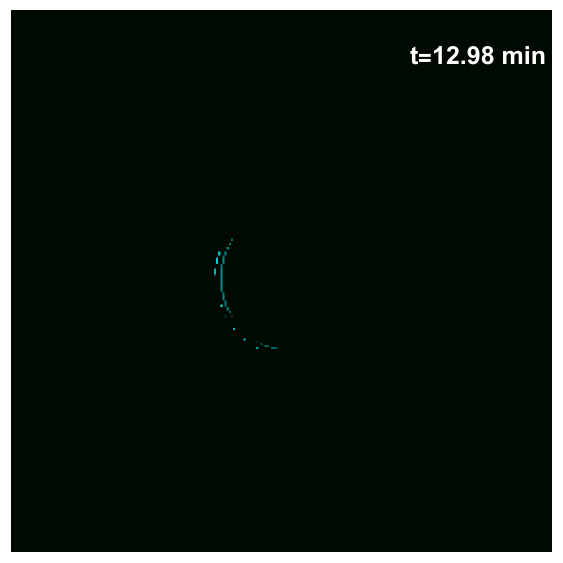} 
  \end{subfigure} 
  \caption{Temporal evolution of the plasma hotspot distribution in the plunging region.}
  \label{fig:2}
\end{figure}

\begin{figure}[!h]
  \centering
  \begin{subfigure}{0.26\textwidth}
    \centering
    \includegraphics[width=\linewidth]{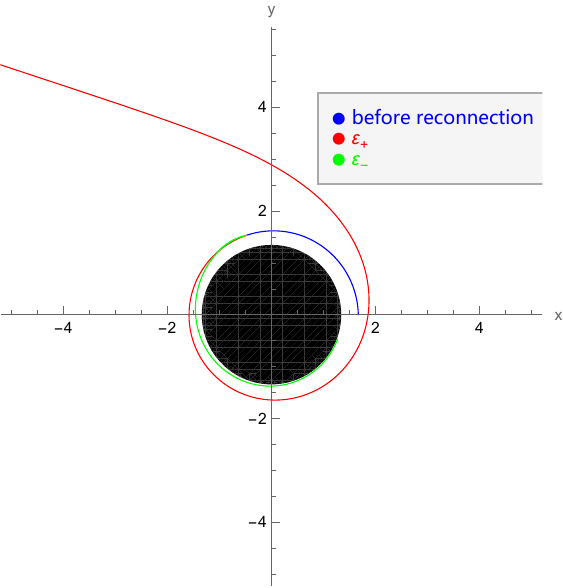}
    \caption{}
    \label{fig:3a}
  \end{subfigure}
  \begin{subfigure}{0.26\textwidth}
    \centering
    \includegraphics[width=\linewidth]{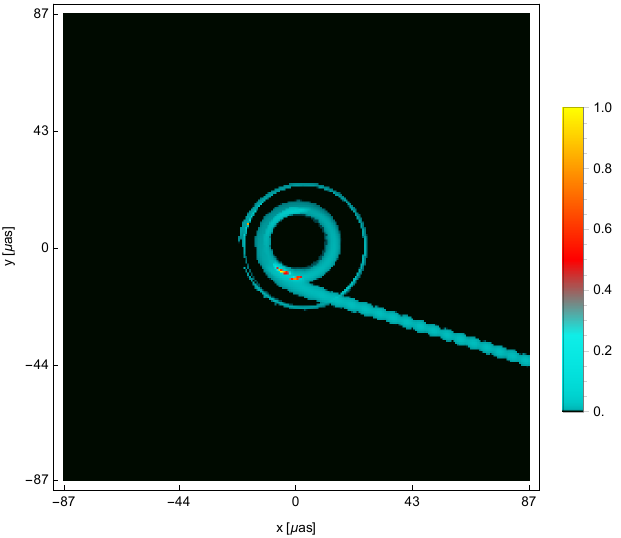}
    \caption{}
    \label{fig:3b}  
  \end{subfigure}
  \begin{subfigure}{0.26\textwidth}
    \centering
    \includegraphics[width=\linewidth]{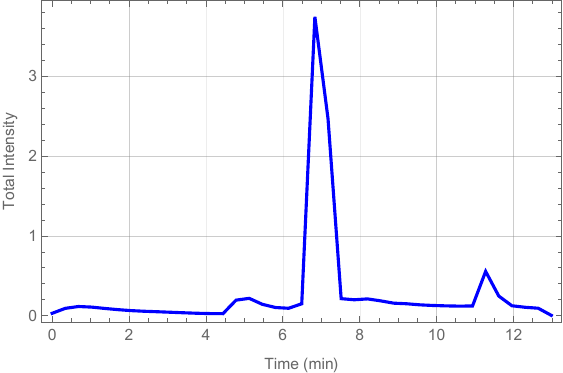}
    \caption{}
    \label{fig:3c}  
  \end{subfigure}
  \begin{subfigure}{0.26\textwidth}
    \centering
    \includegraphics[width=\linewidth]{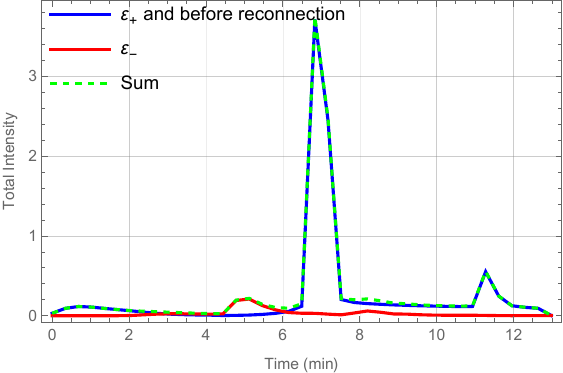} 
    \caption{}
    \label{fig:3d}  
  \end{subfigure}
  \begin{subfigure}{0.26\textwidth}
    \centering
    \includegraphics[width=\linewidth]{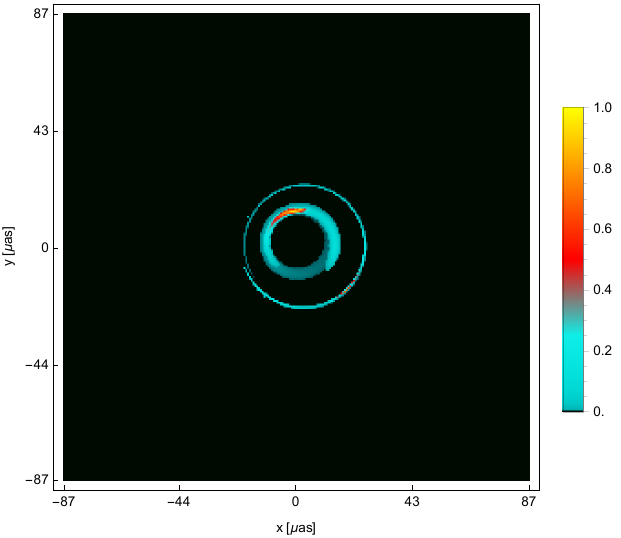} 
    \caption{}
    \label{fig:3e}  
  \end{subfigure}
\caption{In the plunging region: (a) Plasma trajectories in 2D Cartesian coordinates (blue: no reconnection, red: accelerated, green: decelerated, black: black hole). (b) Normalized intensity distribution. (c) Light curve. (d) Light curves: blue solid = $\varepsilon_{+}$ + before reconnection, red solid = only $\varepsilon_{-}$, green dashed = total observed. (e) Normalized intensity from only decelerated plasma.}
  \label{fig:3}
\end{figure}

To clearly observe the post-reconnection process, we take only a short segment before splitting as the no-reconnection hotspot, rather than from ISCO all the way to $r_X$, because the latter would involve many orbits and multiple flares. This follows Ref. \cite{10}, where the Keplerian orbit without reconnection was less than half a circle. We set the initial no-reconnection hotspot at $r=1.67$, $\phi=0$, so the radial range is from $1.67$ inspiraling to $r_X=1.6$. Its proper time is $1.06$, and azimuthal angle traversed is $\phi=1.87$. The portion from ISCO to $r=1.67$ is still inspiraling but not included; this simplification does not affect the analysis of flares from $\varepsilon_{-}$ and $\varepsilon_{+}$. The hotspot imaging for this process is shown in Figs. \ref{fig:2} and \ref{fig:3}.

According to Fig. \ref{fig:2}, at $t=0$ a plunging hotspot appears. At $t=2.05$, the instantaneous magnetic reconnection completes. By $t=3.08$, the accelerated and decelerated plasma hotspots become distinguishable. At $t=4.78$, the hotspot reaches the first flare. At $t=6.83$, it reaches the second flare (the brightest peak); secondary/higher-order images also appear. By $t=8.54$, the accelerated plasma moves outward while the decelerated plasma lingers near the horizon. At $t=11.28$, the hotspot is at the third flare; the decelerated plasma is about to fall in, and the accelerated plasma about to exit the screen. By $t=12.98$, both primary images have left the screen, leaving only secondary/higher-order images. This is similar to Ref. \cite{10}.

The total light curve in Fig. \ref{fig:3c} shows three flares: weak, bright, weak, consistent with Ref. \cite{10} (note that $\varepsilon_{-}$ produces two bumps, the second much fainter; it does not appear as a distinct flare in the total curve). From Fig. \ref{fig:3d}, the first weak flare comes from the decelerated plasma, while the subsequent two come from the accelerated plasma. As explained in Ref. \cite{10}, the bright flare arises because the accelerated plasma gains significant kinetic energy and undergoes strong Doppler blueshift; its secondary images also experience blueshift. The decelerated plasma flare results from Doppler blueshift of its primary image. For details, see Appendix A of Ref. \cite{10}. In Ref. \cite{10}, artificially setting $\varepsilon_{-}>0$ produced almost no flare, showing asymmetry between $\varepsilon_{-}<0$ and $\varepsilon_{-}>0$ plasmas. Ref. \cite{11} also found that $\varepsilon_{-}>0$ produced virtually no flare. Thus, the first flare from $\varepsilon_{-}<0$ may serve as a signature of ongoing Penrose process energy extraction. Our plunging-region hotspot imaging also shows that $\varepsilon_{-}$ produces the first flare, indicating the value of this study. However, observing three flares (or a weak precursor) does not guarantee energy extraction; for example, changing $\xi$ to $\pi/20$ while keeping other parameters still satisfies energy extraction, but $\varepsilon_{-}$ does not produce a flare.

Next, we compare with the circular orbit region under the same parameters. Here $\varepsilon_{-}=-0.33$, $\varepsilon_{+}=8.76$. The no-reconnection Keplerian orbit azimuthal angle is also $1.87$. In the circular region, the current sheet moves at constant radius $r_X$, with energy and angular momentum corresponding to $r_X$ rather than ISCO, and $U^{r}_K=0$. Moreover, the escape condition is always satisfied because $\varepsilon_{+}$ can extend to infinity.

\begin{figure}[!h]
  \centering
  \begin{subfigure}{0.24\textwidth}
    \centering
    \includegraphics[width=\linewidth]{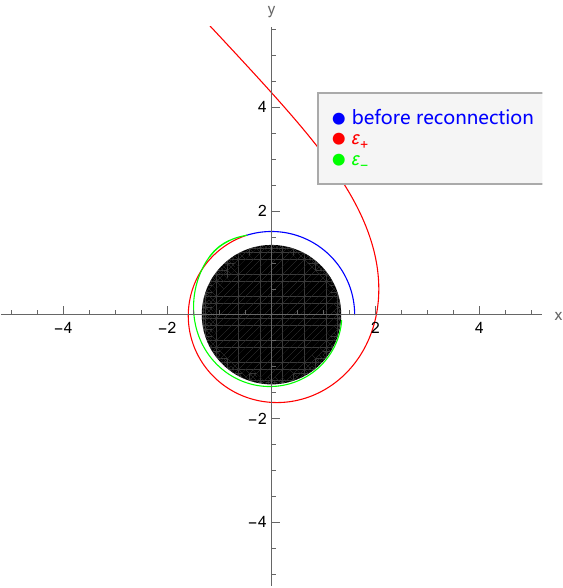}
    \caption{}
    \label{fig:4a}
  \end{subfigure}
  \begin{subfigure}{0.24\textwidth}
    \centering
    \includegraphics[width=\linewidth]{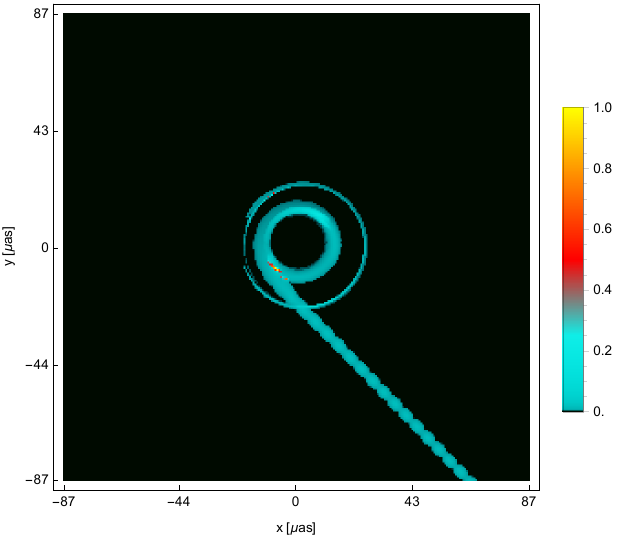}
    \caption{}
    \label{fig:4b}  
  \end{subfigure}
  \begin{subfigure}{0.24\textwidth}
    \centering
    \includegraphics[width=\linewidth]{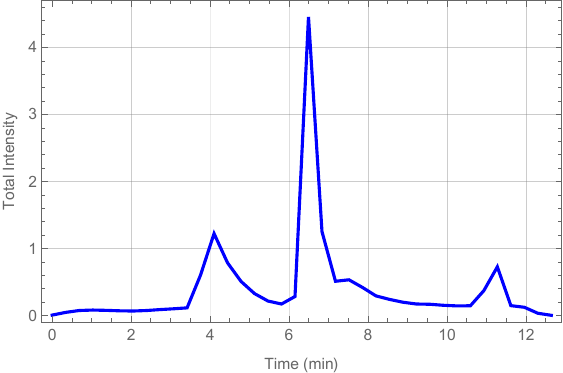}
    \caption{}
    \label{fig:4c}  
  \end{subfigure}
  \begin{subfigure}{0.24\textwidth}
    \centering
    \includegraphics[width=\linewidth]{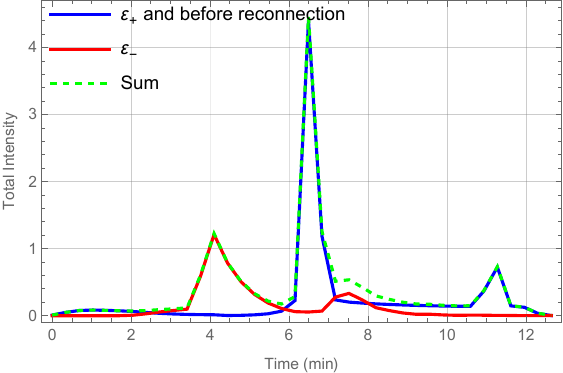} 
    \caption{}
    \label{fig:4d}  
  \end{subfigure}
\caption{In the circular orbit region: (a) Plasma trajectories, (b) Normalized intensity distribution, (c) Light curve, (d) Light curves for three scenarios.}
  \label{fig:4}
\end{figure}

Figure \ref{fig:4} shows that the total light curve still displays three flares, with the first from $\varepsilon_{-}$ and the next two from $\varepsilon_{+}$. Notably, the first flare intensity is significantly stronger than in the plunging region, indicating that the energy extraction signal is more pronounced in the circular orbit region. Moreover, with $\xi=\pi/20$ (still satisfying energy extraction), $\varepsilon_{-}$ still produces a flare in the circular region, whereas in the plunging region it does not. This again demonstrates that the circular region is better for identifying the energy extraction signal.

Now we consider the plunging region when the escape condition is not satisfied. Set $r_X=1.45$ (instead of $1.6$), keeping other conditions unchanged. Only a short segment before splitting is taken as the no-reconnection hotspot.

\begin{figure}[!h]
  \centering
  \begin{subfigure}{0.24\textwidth}
    \centering
    \includegraphics[width=\linewidth]{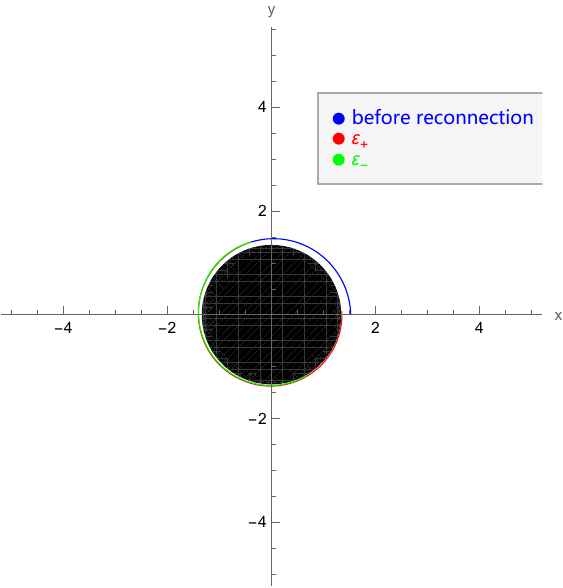}
    \caption{}
    \label{fig:5a}
  \end{subfigure}
  \begin{subfigure}{0.24\textwidth}
    \centering
    \includegraphics[width=\linewidth]{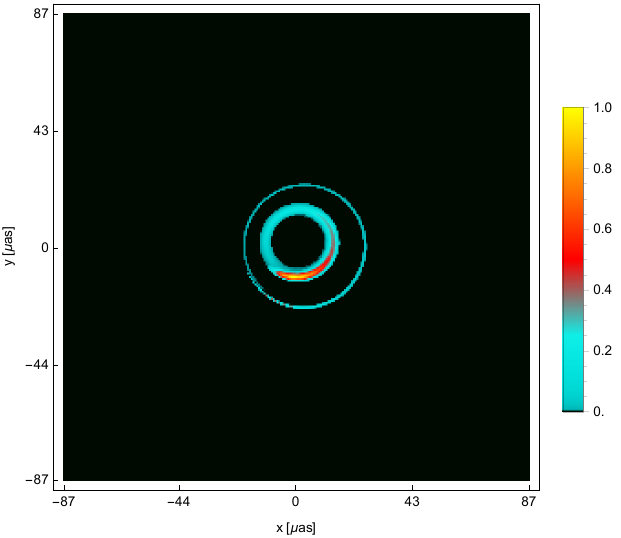}
    \caption{}
    \label{fig:5b}  
  \end{subfigure}
  \begin{subfigure}{0.24\textwidth}
    \centering
    \includegraphics[width=\linewidth]{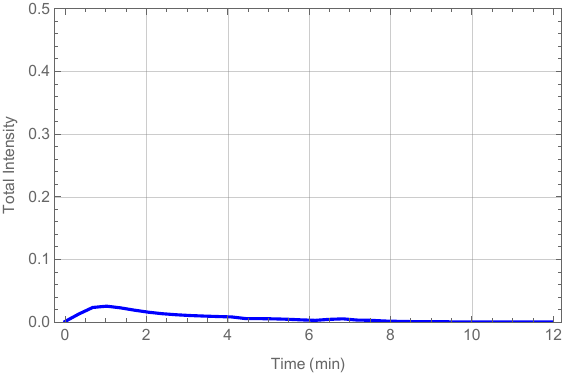}
    \caption{}
    \label{fig:5c}  
  \end{subfigure}
  \begin{subfigure}{0.24\textwidth}
    \centering
    \includegraphics[width=\linewidth]{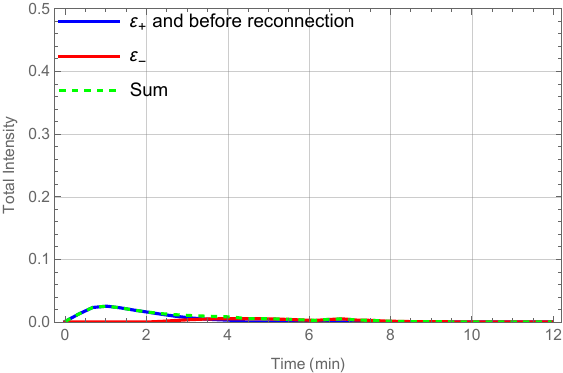} 
    \caption{}
    \label{fig:5d}  
  \end{subfigure}
\caption{Plunging region when escape condition fails: (a) Trajectories, (b) Normalized intensity, (c) Light curve, (d) Light curves for three scenarios.}
  \label{fig:5}
\end{figure}

From Fig. \ref{fig:5}, when $r_X$ is too low and escape condition fails, $\varepsilon_{+}$ does not escape to infinity; the light curve shows no flares, and the total intensity of the plunging current sheet is very weak.

\subsection{$a=0.99$, $r_X=1.3$}

Now we discuss the near-extremal case $a=0.99$, $r_X=1.3$. First, we show no-reconnection hotspot images for both plunging and circular regions. For the circular region, the current sheet moves at constant radius $r=1.3$, completing the same number of orbits as the plunging trajectory.

\begin{figure}[!h]
  \centering
  \begin{subfigure}{0.26\textwidth}
    \centering
    \includegraphics[width=\linewidth]{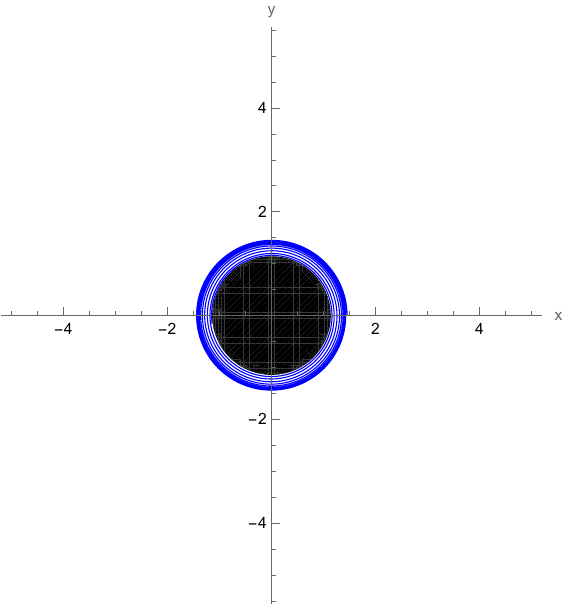}
  \end{subfigure}
  \begin{subfigure}{0.26\textwidth}
    \centering
    \includegraphics[width=\linewidth]{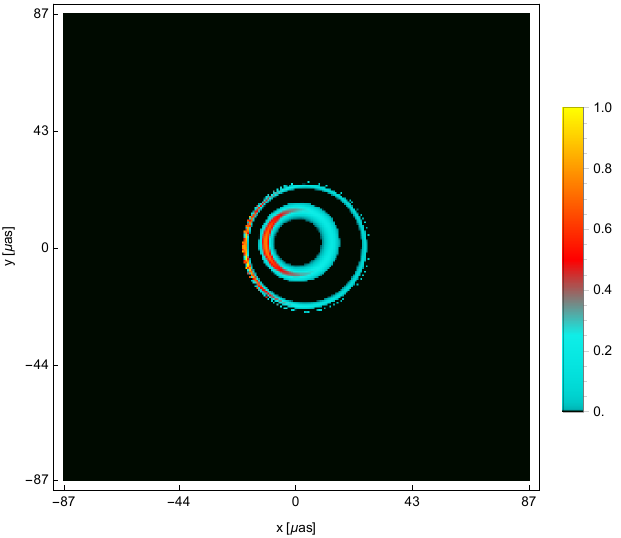}  
  \end{subfigure}
  \begin{subfigure}{0.26\textwidth}
    \centering
    \includegraphics[width=\linewidth]{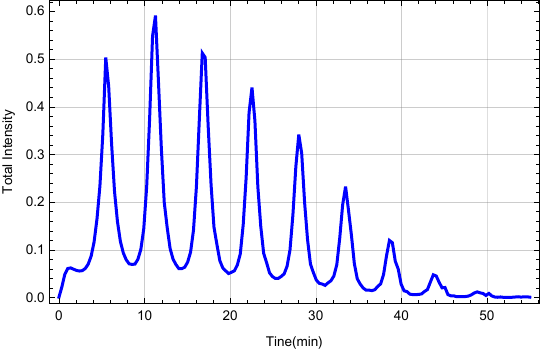} 
  \end{subfigure}
  \begin{subfigure}{0.26\textwidth}
    \centering
    \includegraphics[width=\linewidth]{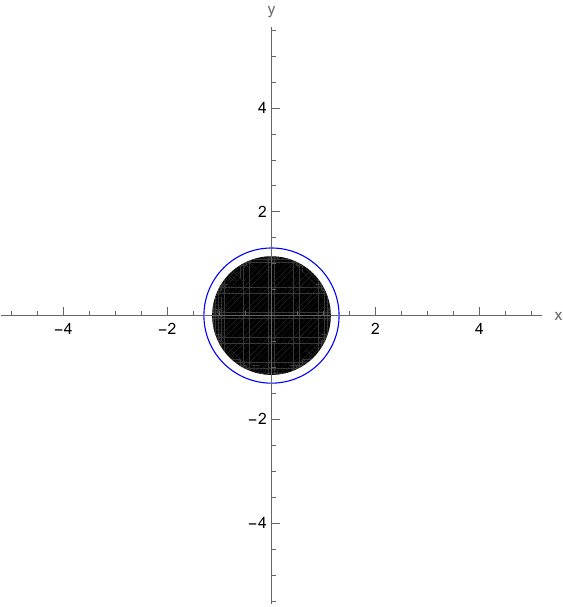}   
  \end{subfigure}
  \begin{subfigure}{0.26\textwidth}
    \centering
    \includegraphics[width=\linewidth]{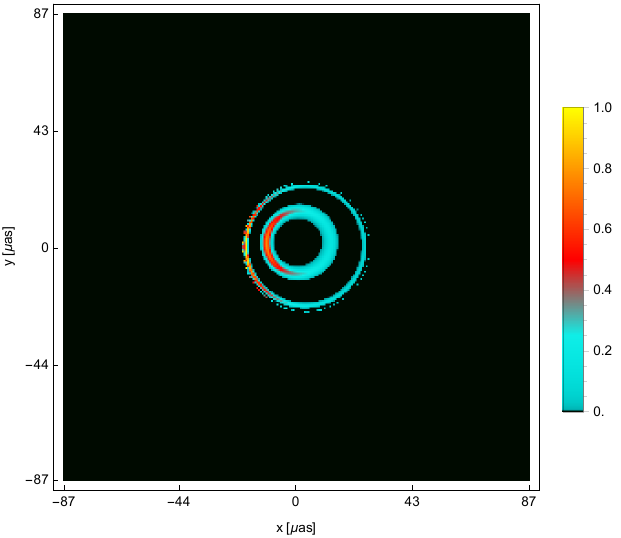}   
  \end{subfigure}
   \begin{subfigure}{0.26\textwidth}
    \centering
    \includegraphics[width=\linewidth]{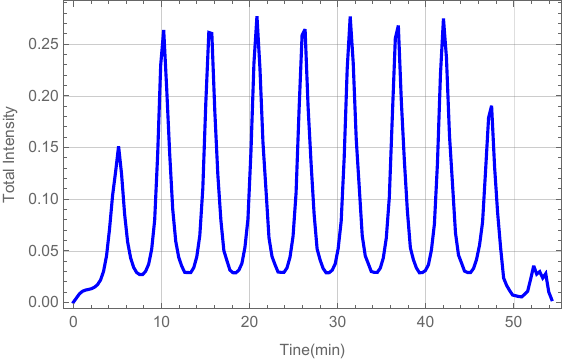}   
  \end{subfigure} 
\caption{For $a=0.99$: upper row = plunging region without reconnection; lower row = circular region without reconnection. First column: trajectories, second column: normalized intensity, third column: light curves.}
  \label{fig:6}
\end{figure}

From Fig. \ref{fig:6}, in the plunging region we see progressively weakening flares, consistent with the $a=0.94$ case. In the circular region, flares of roughly constant intensity appear, analogous to Ref. \cite{10}. This reaffirms the accuracy of our plunging-region hotspot imaging.

Next, we present hotspot images after magnetic reconnection. The no-reconnection hotspot only covers a short segment before splitting, with the same azimuthal coverage. Both scenarios satisfy energy extraction. We plot results for both plunging and circular regions.

\begin{figure}[!h]
  \centering
  \begin{subfigure}{0.24\textwidth}
    \centering
    \includegraphics[width=\linewidth]{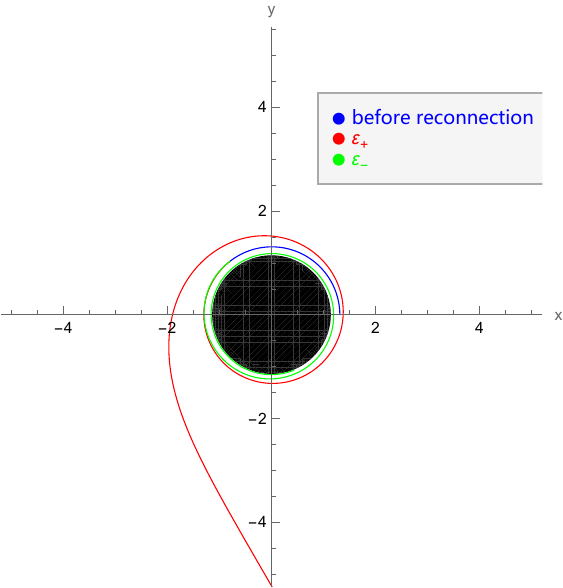}
  \end{subfigure}
  \begin{subfigure}{0.24\textwidth}
    \centering
    \includegraphics[width=\linewidth]{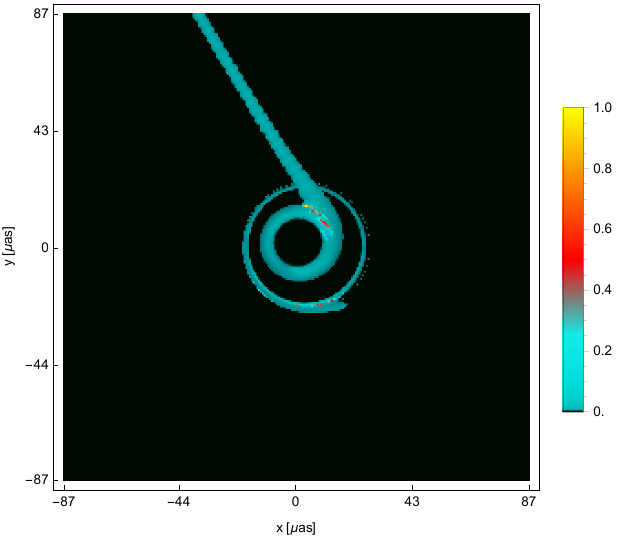}  
  \end{subfigure}
  \begin{subfigure}{0.24\textwidth}
    \centering
    \includegraphics[width=\linewidth]{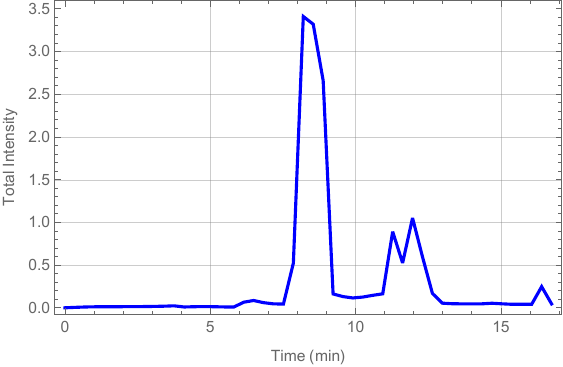} 
  \end{subfigure}
  \begin{subfigure}{0.24\textwidth}
    \centering
    \includegraphics[width=\linewidth]{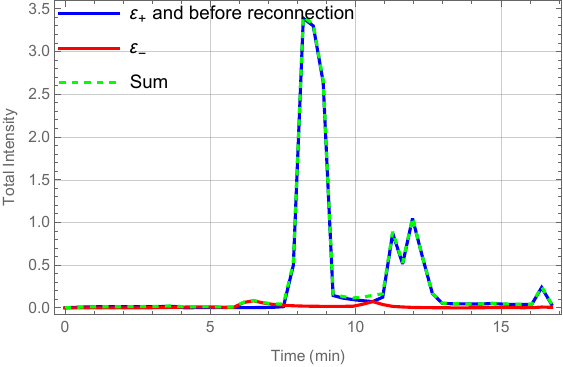}   
  \end{subfigure}
  \begin{subfigure}{0.24\textwidth}
    \centering
    \includegraphics[width=\linewidth]{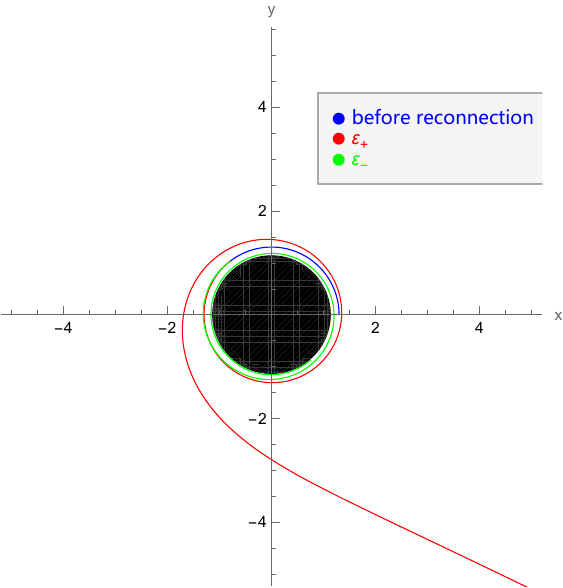}   
  \end{subfigure}
   \begin{subfigure}{0.24\textwidth}
    \centering
    \includegraphics[width=\linewidth]{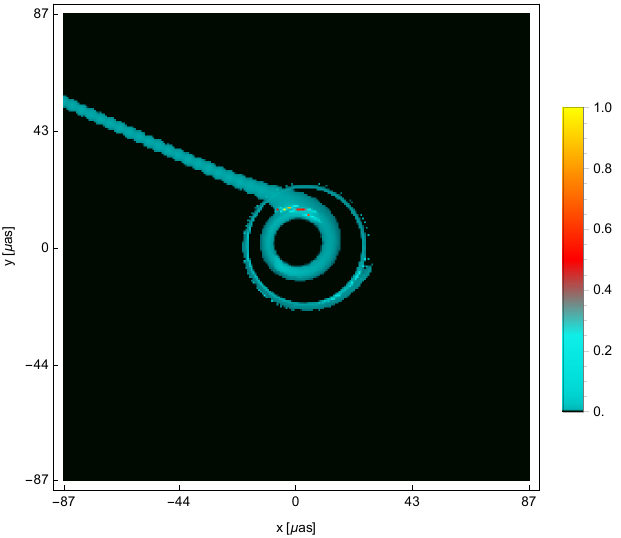} 
  \end{subfigure}
  \begin{subfigure}{0.24\textwidth}
    \centering
    \includegraphics[width=\linewidth]{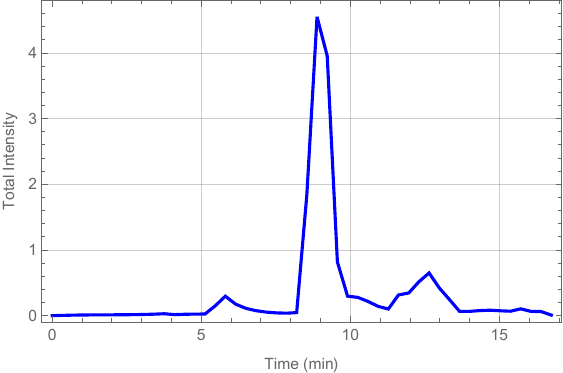}   
  \end{subfigure}
  \begin{subfigure}{0.24\textwidth}
    \centering
    \includegraphics[width=\linewidth]{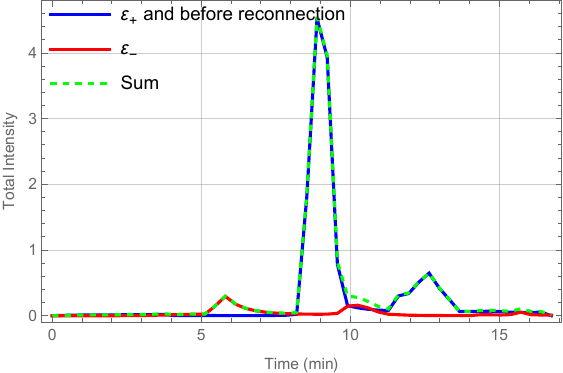}   
  \end{subfigure}
\caption{For $a=0.99$ with magnetic reconnection: upper row = plunging region, lower row = circular region. First column: trajectories, second column: normalized intensity, third column: light curves, fourth column: light curves for three scenarios.}
  \label{fig:7}
\end{figure}

From the upper row of Fig. \ref{fig:7}, in the plunging region we observe four flares within the observation time, differing from previous findings. The fourth column shows that all four flares originate from $\varepsilon_{+}$. The first flare is bright, the second and third are relatively weak and close in time, and the last is very faint. Before the first bright flare, $\varepsilon_{-}$ produces a slight bump too weak to be counted as a flare. Thus, for near-extremal black holes, identifying energy extraction becomes more difficult, consistent with Refs. \cite{10,11}. From the lower row, in the circular region we observe three flares, with the first from $\varepsilon_{-}$ and the next two from $\varepsilon_{+}$. Compared to the plunging region where $\varepsilon_{-}$ does not produce a flare, this again shows that the circular region is more effective for identifying energy extraction. Comparing Fig. \ref{fig:4} (a=0.94) with the lower row of Fig. \ref{fig:7} (a=0.99), the first flare for a=0.99 is much weaker, confirming that higher spin makes energy extraction identification more challenging, in agreement with Ref. \cite{11}\footnote{Although the $r_X$ values differ between the two figures, we have verified that even when $r_X$ is the same, the first flare for $a=0.99$ is still weaker than that for $a=0.94$.}.

\section{Conclusion}

In this paper, we employed the hotspot imaging method to observe the process before and after the Comisso-Asenjo mechanism in the plunging region of a Kerr black hole. We first reviewed the magnetic reconnection process in the plunging region, then introduced the hotspot model and imaging technique, and finally presented numerical observational results. Our analysis focused on two scenarios: $a=0.94, r_X=1.6$ and $a=0.99, r_X=1.3$. For the first case, we plotted no-reconnection hotspot images, post-reconnection images, and images where $\varepsilon_{+}$ fails to escape; we also compared with the circular orbit region. For the second case, we similarly plotted no-reconnection and post-reconnection images, and compared with the circular region. We found that for no-reconnection hotspot images, when the plasma follows plunging orbits, observers see flares with gradually decreasing intensity; for circular orbits, flares have nearly constant intensity. In the plunging region, the energy extraction signal is less pronounced than in the circular orbit region. Therefore, to identify the energy extraction signal, it is preferable to conduct observations in the circular orbit region.

Beyond purely analytical approaches, numerical simulations are essential for a deep understanding of magnetic reconnection and associated phenomena such as turbulence.

\noindent {\bf Acknowledgments}

\noindent
This work is supported by the National Natural Science Foundation of China (Grants Nos.
12375043, 12575069 ), and Chongqing Normal University Fund Project (Grants No. 26XLB001).

\end{document}